\documentclass[a4paper,11pt]{article}

\usepackage[T1]{fontenc} 
\usepackage{color}
\usepackage{ulem}
\usepackage{braket}
\usepackage{mathtools}
\usepackage{graphicx,subfigure,adjustbox}
\usepackage{comment}

\usepackage{jheppub} 

\preprint{RESCEU-24/21, RIKEN-iTHEMS-Report-21}

\title
{Vacuum decay in the Lorentzian path integral}
\author[a,b]{Takumi Hayashi,}
\author[b]{Kohei Kamada,}
\author[c]{Naritaka Oshita,}
\author[a,b,d,e]{Jun'ichi Yokoyama}

\affiliation[a]{Department of Physics, Graduate School of Science,
The University of Tokyo,\\ Hongo 7-3-1
Bunkyo-ku, Tokyo 113-0033, Japan}
\affiliation[b]{Research Center for the Early Universe (RESCEU), Graduate School of Science,\\ The University of Tokyo, Hongo 7-3-1
Bunkyo-ku, Tokyo 113-0033, Japan}
\affiliation[c]{RIKEN iTHEMS, Wako, Saitama 351-0198, Japan}
\affiliation[d]{Kavli Institute for the Physics and Mathematics of the Universe (Kavli
 IPMU), UTIAS, WPI, The University of Tokyo, Kashiwa, Chiba, 277-8568, Japan}
\affiliation[e]{Trans-scale Quantum Science Institute,\\ The University of Tokyo, Hongo 7-3-1
Bunkyo-ku, Tokyo 113-0033, Japan}
\emailAdd{takumi\_hayashi@resceu.s.u-tokyo.ac.jp}
\emailAdd{kohei.kamada@resceu.s.u-tokyo.ac.jp}
\emailAdd{naritaka.oshita@riken.jp}
\emailAdd{yokoyama@resceu.s.u-tokyo.ac.jp}

\begin{document}

\abstract{We apply the Lorentzian path integral to the decay of a false vacuum and estimate the false-vacuum decay rate. To make the Lorentzian path integral convergent, the deformation of an integral contour is performed by following the Picard-Lefschetz theory. We show that the nucleation rate of a critical bubble, for which the corresponding bounce action is extremized, has the same exponent as the Euclidean approach. We also extend our computation to the nucleation of a bubble larger or smaller than the critical one to which the Euclidean formalism is not applicable.}

\maketitle

\section{Introduction}
The first order phase transition of false vacuum (or false vacuum decay) has been studied as a non-perturbative phenomenon in quantum field theory \cite{Coleman:1977py,Callan:1977pt,Coleman:1980aw}. This is an important quantum process in cosmological scenarios such as the old inflation \cite{Sato:1980yn,Guth:1980zm} and open inflation~\cite{Sasaki:1993ha,Yamamoto:1995sw,Bucher:1994gb}. Recently,  study of false vacuum decay is highly motivated in the context of the meta-stability of the Standard Model Higgs potential, too \cite{Chatrchyan:2012ufa,Sher:1988mj,Arnold:1989cb,Altarelli:1994rb,Espinosa:1995se,Casas:1996aq,Hambye:1996wb,Isidori:2001bm,Espinosa:2007qp,Ellis:2009tp,Bezrukov:2012sa,Bednyakov:2015sca,EliasMiro:2011aa,Degrassi:2012ry,Buttazzo:2013uya,Branchina:2013jra,Greenwood:2008qp,Chigusa:2017dux,Chigusa:2018uuj}. Although there are many interesting applications to cosmological or particle physics scenarios there are yet critical issues in the fundamental theory of false vacuum decay such as the symmetry of the bounce solution \cite{Coleman:1977py,Coleman:1977th} and the negative mode problem \cite{Lavrelashvili:1985vn,Tanaka:1992zw,Lavrelashvili:1999sr,Tanaka:1999pj,Khvedelidze:2000cp,Gratton:2000fj,Dunne:2006bt,Lee:2014uza,Koehn:2015hga,Gregory:2018bdt,Bramberger:2019mkv}. Another major issue is that the computation of vacuum decay rate with gravity relies on the Euclidean path integral without justification \cite{Hebecker:2018ofv}.
The Euclidean path integral for a gravitational system is not bounded below and unstable for conformal perturbations. However, most of the studies applying the vacuum decay process to cosmological situations rely on the Euclidean computation developed by Coleman and de Luccia \cite{Coleman:1980aw}. 

The Euclidean formalism in gravity was originally developed in the context of quantum cosmology. It leads to a number of interesting proposals such as creation of an inflationary universe from nothing \cite{Vilenkin:1982de} and the no-boundary proposal \cite{Hawking:1981gb,Hawking:1983hj}. However, They largely obtain different conclusions with different behaviors of the wave function of the universe.
One of the origin of these discrepancies lies in the ambiguity of the direction of the Wick rotation \cite{Linde:1983cm} or which saddle configuration to incorporate. 
Thus it is desirable if we can perform the path integral in the Lorentzian spacetime, 
which is difficult in practice.   
Recently, however, the Lorentzian path integral 
of gravitational action was successfully performed using the Picard-Lefschetz theory
albeit in mini superspace~\cite{Feldbrugge:2017kzv}. 
This computational technique has a significant advantage since it tells us which saddle points should be taken or not,  and we do not need to assume a priori the Euclidean path integral contour i.e., a Wick rotated contour $t\to -it$.
The Lorentzian path integral is thus useful to clarify the ambiguities in Euclidean formalism and has also been applied to a number of problems~\cite{Turok:2013dfa,Baldazzi:2019kim,Feldbrugge:2019fjs,Rajeev:2021yyl,Rajeev:2021zae}.

In this paper, we perform the Lorentzian path integral for the vacuum decay process
with applying the Picard-Lefschetz theory, 
re-estimate the decay rate of a false vacuum, and compare it with the result obtained in the Euclidean formalism. 
The Lorentzian path integral is difficult to perform even with  the 
Picard-Lefschetz theory because of the infinite dimensional space of the theory. 
Here we describe a system with a bubble in terms of the Polyakov action, 
in a similar way to Ref.~\cite{Basu:1991ig}, 
which reduces the degree of freedom of the system, so that we can evaluate the Lorentzian 
path integral. Although our final goal is to formulate a theory of vacuum decay with gravity in terms of the Lorentzian path integral, we here consider situations where gravitational backreaction can be ignored, as the first step. Even for a vacuum decay without gravity, one can see some advantages to perform  Lorentzian path integral for the vacuum decay process. In the Euclidean formalism, one can estimate the nucleation rate of the critical bubble for which the Euclidean action is minimized. On the other hand, the Lorentzian technique enables us to handle even large or small-bubble nucleation, for which the Euclidean action is not minimized.

The rest of the paper is organized as follows. In Sec. \ref{sec_euclidean}, we briefly review the computation of vacuum decay rate in the Euclidean path integral. In Sec. \ref{sec_Lorentzian}, we perform a Lorentzian path integral to compute the nucleation rate of a vacuum bubble whose dynamics is governed by the Polyakov-type action. The final section is dedicated to conclusions and  interpretation of the result of the Lorentzian path integral. In the Appendix, we briefly review the Picard-Lefschetz theory, by which one can find an integral contour that makes the integral absolutely convergent.

\section{Vacuum decay in the Euclidean path integral}
\label{sec_euclidean}
In this section, we briefly review the vacuum decay in the de Sitter and Minkowski spacetime using the Euclidean path integral. 
We start with the standard procedure developed by Coleman and de Luccia~\cite{Coleman:1980aw}, including the analytic evaluation with the thin-wall approximation. 
We then introduce the  effective theory approach of the bubble walls with the Nambu-Goto action~\cite{Basu:1991ig} (see also Ref.~\cite{Ai:2020vhx}), which is classically equivalent to the Polyakov action, 
and show its equivalence to the standard approach in the bubble nucleation rate.
We will see that the Lorentzian path integral can be evaluated 
with the help of the Picard-Lefschetz theory in the next section. 

The 4-dimensional de Sitter space is given as a hyperboloid
\begin{align}
-{\tilde T}^2+{\tilde X}^2+{\tilde Y}^2+{\tilde Z}^2+{\tilde W}^2=l^2, \label{5d4d}
\end{align}
embedded in the 5-dimensional Minkowski spacetime with the coordinate
$({\tilde T}, {\tilde X}, {\tilde Y}, {\tilde Z}, {\tilde W})$
where $l$ is the curvature scale of the 4-dimensional de Sitter spacetime. We take the slicing of ${\tilde X}=l\cos\frac\xi l$ that leads to the foliation on which the line element is \cite{Coleman:1980aw,Leblond:2002ns}
\begin{align}
ds_{\mathrm{dS_4}}^2 &= d\xi^2+\rho(\xi)^2 ds^2_\mathrm{dS_3} \ \ \text{with} \ \rho(\xi) \equiv l\sin \frac\xi l,
\label{de_metric}\\
ds_\mathrm{dS_3}^2 &= -d\eta^2+ \cosh^2\eta(d\theta^2+\sin^2\theta d\phi^2),\label{de_metric2}
\end{align}
where $\xi \in [0, \pi l]$, 
so that the metric respects the $O(4,1)$ symmetry. 
By performing the Wick rotation for $\eta$, it leads to the $O(5)$ symmetric Euclidean de Sitter space.
Note that the chart $(\xi,\eta,\theta,\phi)$ covers a part of de Sitter spacetime, $|{\tilde X}|<l$. In the flat limit ($l\to \infty$), the function $\rho (\xi)$ reduces to $\rho=\xi$, which reproduce the Rindler-type coordinate of Minkowski spacetime.

To investigate the false vacuum decay, we consider the action of a scalar field $\varphi$
\begin{align}
S[\varphi]=\int d^4x\sqrt{-\det g_{\mu\nu}}\left(-\frac 12 g^{\mu\nu} \partial_\mu\varphi \partial_\nu \varphi-V(\varphi)\right)
\end{align}
where $V(\varphi)$ has a metastable false vacuum at $\varphi =\varphi_\mathrm{fv}$ and a stable true vacuum at $\varphi=\varphi_\mathrm{tv}$. 
The transition probability $P$ is estimated by the Euclidean bounce action $B$ as~\cite{Callan:1977pt}
\begin{align}
\begin{split}
&P\sim e^{-B},\quad \mathrm{with}\  B=S_E[\varphi_b]-S_E[\varphi_\mathrm{fv}],\\
&S_E[\varphi]=\int d^4x_\mathrm{E}\sqrt{\det g_{\mathrm{E}\mu\nu}}\left[\frac 12 g^{\mu \nu}_{\rm E} \left(\frac{\partial \varphi}{\partial x_{\rm E}^{\mu}} \right) \left(\frac{\partial \varphi}{\partial x_{\rm E}^{\nu}} \right)+V(\varphi)\right],
\end{split}
\end{align}
where $S_E$ is the Euclidean action, and $x_{\rm E}^{\mu}$ and $g_{\rm E}$ is the Euclidean coordinate and metric, respectively. $\varphi_b$ is a non-trivial Euclidean bounce solution (or the ``bubble'' configuration) that connects the false vacuum 
and the true vacuum. 
We here neglect the backreaction of the bubble configuration, to the background spacetime. 
The minimum bounce action describing the most probable process of vacuum decay is realized, at least in the absence of gravity  \cite{Coleman:1977th}, by a maximally symmetric ($O(4)$ symmetric) bounce solution 
that depends only on $\xi$. The Lorentzian metric described by Eqs.~\eqref{de_metric} and \eqref{de_metric2} is 
analytically continued to
the Euclidean metric of the $4$-sphere by taking $\eta\to-i(\eta_\mathrm{E}-\pi/2)$ as,
\begin{align}
ds_{\mathrm{S_4}}^2= d\xi^2+\rho(\xi)^2 ds_\mathrm{S_3}^2,\quad \mathrm{with}\ ds_\mathrm{S_3}^2 = d\eta_\mathrm{E}^2+ \sin^2\eta_\mathrm{E}(d\theta^2+\sin^2\theta d\phi^2), 
\end{align}
where $\rho (\xi)$ is fixed to $l \sin{(\xi/l)}$, which is appropriate to 
find the desirable bounce solution.

The bounce equation for $\varphi(\xi)$ is now given as
\begin{align}
\partial_{\xi}^2\varphi+3\frac{\partial_{\xi}\rho}\rho\partial_{\xi}\varphi-\frac{dV}{d\varphi}=0. \label{bounce_eq}
\end{align}
As the boundary conditions, we require
\begin{align}
&\partial_{\xi}\varphi(\xi)\sim \varphi_\mathrm{tv}, \quad \varphi(\xi)\to 0, \quad(\xi\to 0),\\
&\partial_{\xi}\varphi(\xi)\sim \varphi_\mathrm{fv}, \quad \varphi(\xi)\to 0, \quad(\xi\to \pi l), 
\label{regularity_south}
\end{align}
so that the scalar field approaches to true and false vacua at the North ($\xi =0$) 
and South ($\xi = \pi l$) poles, respectively. 
With these conditions we avoid the singularities  there. 
For the case of the flat Euclidean space ($l\to\infty$), we do not need the condition of regularity of $\varphi$ at infinity.

The Lorentzian evolution of the nucleated bubble on the chart $|{\tilde X}|<l$ is easily obtained by $\varphi(x) = \varphi_b(\xi)$ with an inverse Wick rotation $(\eta_E-\pi/2)\to i\eta$, 
but it does not cover the whole 4-dimensional de Sitter spacetime Eq.~\eqref{5d4d}. 
The evolution in the rest of the spacetime is obtained by another analytic continuation $\xi\to i\tilde \xi, \eta_\mathrm{E}\to i\tilde\eta$, which gives the open slicing chart $\tilde x$ of the de Sitter spacetime in the region, $|{\tilde X}|>l$, with the metric
\begin{align}
ds_{\mathrm{dS_4}}^2=-d\tilde\xi^2 + l^2\sinh^2\frac{\tilde\xi}{l} \left[ d\tilde\eta^2+ \sinh^2\tilde\eta(d\theta^2+\sin^2\theta d\varphi^2)\right].
\end{align}
The Lorentzian solution on the chart  $|{\tilde X}|>l$ is then obtained by taking $\varphi(\tilde x)=\varphi_b(i\tilde\xi)$. 
In such a way, the dynamics of the bubble after nucleation
is described in the Lorentzian 4-dimensional de Sitter spacetime Eq.~\eqref{5d4d}.
Thanks to the regularity of the bounce solution at $\xi=0$, the solutions in the two charts 
are smoothly connected. Note that the parallel argument is valid for the case of Minkowski spacetime.

Although it is necessary to solve the non-trivial bounce equation~\eqref{bounce_eq} to evaluate the vacuum decay rate, exact solutions are difficult to  obtain analytically. On the other hand,
the thin wall approximation (if applicable) allows us to evaluate the bounce action analytically. It is valid when the bubble configuration exhibits a distinct interior, where the scalar field sets around the true vacuum, and the wall,  
where the value of $\varphi$ changes drastically from the true vacuum to the false vacuum, and the width of bubble wall, $\Delta \xi$, is much smaller than the bubble radius $\xi_b$. Then, the Euclidean action can be divided into the boundary (wall) and bulk (interior) parts. In the former part, we can neglect the second term in~\eqref{bounce_eq} so that ${(\partial_{\xi} \varphi)}^2 \sim 2 (V(\varphi)-V(\varphi_\mathrm{fv}))$, which yields
\begin{align}
\begin{split}
B&= \int d\Omega_3\left[\rho(\xi_b)^3 \int_{\xi_b}^{\xi_b+\Delta\xi} d\xi (\partial_{\xi}\varphi_b)\sqrt{2(V(\varphi_b)-V(\varphi_\mathrm{fv}))} + \int_{0}^{\xi_b} d\xi\rho^3(\xi) (V(\varphi_\mathrm{tv})-V(\varphi_\mathrm{fv}))\right]. 
\end{split}
\end{align}
By defining the bubble tension $\sigma$ and the physical bubble radius $\rho_b$ as
\begin{align}
\sigma \equiv \int _{\varphi_{\mathrm{fv}}}^{\varphi_{\mathrm{tv}}}d\varphi\sqrt{2(V(\varphi)-V(\varphi_\mathrm{fv}))}, \quad \rho_b\equiv \rho(\xi_b), \quad \text{and also} \quad \Delta\! V \equiv V(\varphi_\mathrm{fv})-V(\varphi_\mathrm{tv}),
\end{align}
the bounce action can be rewritten in a simpler form as
\begin{align}
\begin{split}
B&= 2\pi^2 \left( \sigma\rho_b ^3 -\int_0^{\xi_b} d\xi \rho^3 (\xi) \Delta\! V\right).
\end{split}
\label{bounce_action_thin}
\end{align}
The bounce is a saddle point solution of the bounce action, and the value of $\xi_b$ can be determined by minimizing the Euclidean action \eqref{bounce_action_thin} with respect to $\xi_b$, which yields
\begin{align}
\rho_b = \partial_{\xi}\rho|_{\xi=\xi_b}\rho_0,\quad \mathrm{with} \ \rho_0=\frac{3\sigma }{\Delta\! V}.\label{radius_condition}
\end{align}
For the case of Minkowski spacetime, where $\rho(\xi) = \xi$, we obtain the bubble radius and the bounce action as~\cite{Coleman:1977py}
\begin{align}
\rho_b=\rho_0,\quad B= \frac{27\pi^2\sigma^4}{2\Delta\! V^3}.\label{bounce_flat}
\end{align}
For the de Sitter spacetime where $\rho = l \sin (\xi/l)$, one notices that 
the positive brunch, $\dot\rho=+\sqrt{1-\rho^2/l^2}$, should be taken from Eq.~\eqref{radius_condition}. Then the physical bubble radius is determined as
\begin{align}
\rho_b = \frac1{\sqrt{1/\rho_0^{2}+1/l^{2}}},\label{bubble_radius}
\end{align}
and the bounce action~\eqref{bounce_action_thin} is rewritten as
\begin{align}
\begin{split}
B&= 2\pi^2 \left( \sigma\rho_b ^3 -\int_0^{\rho_b} d\rho \frac{\rho^3}{\sqrt{1-\rho^2/l^2}} \Delta\! V\right),\\
&=2\pi^2\sigma\left\{\rho_b^3-\frac{l^4}{\rho_0}\left[2-\sqrt{1-{\rho_b^2}/{l^2}}\left(2+{\rho_b^2}/{l^2}\right)\right]\right\} = \frac{2 \pi^2 \sigma \rho_b^3}{(1+\rho_b/\rho_0)^2}.
\end{split}
\label{bounce_action_result1}
\end{align}
It reduces to \eqref{bounce_flat} in the flat limit, $l\to \infty$ which leads to $\rho_b=\rho_0$. If we take another limit $\Delta\!V\to 0$, equivalently $\rho_0\to\infty$, the two vacua degenerate and the simple result is obtained as
\begin{align}
\rho_b=l,\quad B&= 2\pi^2 \sigma l^3.
\end{align}

The formulation described above is a well-known standard procedure, 
but it is difficult to apply it directly to the Lorentzian formalism.
For this purpose, we here introduce a one-dimensional simplified formalism that reproduces the same results of the one 
derived in the above with the thin-wall approximation,
where the degrees of freedom are given only by the trajectory of the bubble wall and other details 
of the system are a priori coarse-grained into $\sigma$ and $\Delta V$, in a similar way to Ref.~\cite{Basu:1991ig}. We will also give a little generalization to the case of non-degenerate vacua. 
Let us consider the static chart of the (bulk) de Sitter spacetime $\left\{ \tilde {x}^{\mu} \right\}=\left\{ t,r,\theta,\phi \right\}$ with the metric
\begin{align}
g_{\mu\nu}d\tilde x^\mu d\tilde x^\nu=-f{dt}^2+f^{-1}dr^2+r^2(d\theta^2 +\sin\theta^2d\phi^2),\quad f(r)=1 -\frac{r^2}{l^2}.
\label{ds_metric_static}
\end{align}
where the limit of $l\to \infty$ reproduces the spherical coordinate system of Minkowski spacetime. This choice of metric is useful for reformulating the problem in a one-dimension model 
since the metric depends only on the radial coordinate.
The thin-wall approximation allows us to describe
the dynamics of bubble by the world volume theory embedded into the background spacetime (\ref{ds_metric_static}) with
the Nambu-Goto (NG) type action as
\begin{align}
S_{\mathrm{NG}}[X^\mu]= -\sigma\int_{\cal W} d^3x \sqrt{-\mathrm{det}h_{ab}}+\Delta\!V\int_{\cal B} d^4\tilde x\sqrt{-\det g_{\mu\nu}},
\quad \mathrm{with}\  h_{ab} =\partial_a X^\mu \partial_b X^\nu g_{\mu\nu},\label{NGaction_wall}
\end{align}
where $\{X^\mu(x)\}=\{T(x),R(x),\Theta(x),\Phi(x)\}$ is the trajectory of the bubble wall embedded in the bulk spacetime, $\{x^a\}=\{\tau,\sigma_1,\sigma_2\}$ is the 3-dimensional coordinate covering the bubble wall, $\cal W$, and $h_{ab}$ is the induced metric on the bubble. In the second term the bulk integration is performed over the bubble interior, denoted by $\cal B$. In the following, we assume the spherical symmetry of the vacuum bubble parametrized by $\{x^a\}=\{\tau,\theta,\phi\}$, for which the trajectory of the bubble is characterized only by the temporal coordinate $\tau$, and the trajectory has the form of $\{X^\mu(x)\}=\{T(\tau),R(\tau),\theta,\phi\}$.  Fixing the gauge as $\tau=t$, the induced metric is given by
\begin{align}
h_{ab}dx^adx^b= \left[-f(R)+f(R)^{-1}\left(\frac{dR}{dt}\right)^2\right] dt^2+R^2 (d\theta^2+\sin^2\theta d\phi^2),\label{induced_metric}
\end{align}
Then the system can be described by the action for the bubble radius $R$ as
\begin{align}
S_{\mathrm{NG}}[R]=4\pi\sigma\int dt \left[-R^2\sqrt{f(R)-f(R)^{-1}\left(\frac{dR}{dt}\right)^2}+ \frac{R^3}{\rho_0}\right],\label{NGaction_R}
\end{align}
where we have used Eq.~\eqref{radius_condition} for the bulk component (second term). 

We now consider the nucleation of the wall from the {\it nothing}, i.e., $R=0$ to $R=\rho_b$. 
From the action Eq.~\eqref{NGaction_R},  one obtains the integrated equation of motion as
\begin{align}
E= 4\pi\sigma\left(\frac{fR^2}{\sqrt{f-f^{-1}\left(\frac{dR}{dt}\right)^2}}-\frac {R^3} {\rho_0}\right),\label{bubble_energy}
\end{align}
where $E$ is the conserved energy (an integration constant), which one can rewrite as
\begin{align}
\left(\frac{dR}{dt}\right)^2 + 2U(R;E)=0, \quad \mathrm{with} \quad  2U(R;E)= f^2 \left[ \frac{R^{4}f}{\left(E/4\pi\sigma+R^3/\rho_0\right)^2}-1\right]. \label{integratedEOM_NG}
\end{align}
Note that $U(R;0)=0$ is satisfied for $R=l$ and $\rho_b$ (Eq.~\eqref{bubble_radius}).

\begin{figure}
    \centering
    \includegraphics[width=10cm]{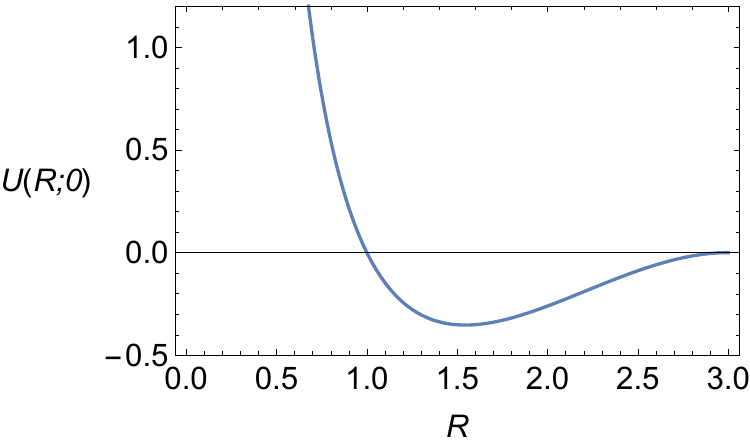}
    \caption{The effective potential of the bubble wall $U(R;0)$ with $l=3$ and $\rho_0=3/(2\sqrt2 )$ is shown. The potential is positive for $R<\rho_b=1$ and the classical dynamics is not allowed in that region. 
    Classically the nucleated bubble at $R=\rho_b$ expands and runs away up to the de Sitter radius but does not shrink.}
    \label{fig:potential}
\end{figure}

Since the bubble nucleation from the state $R=0$ to $R \not =0$ is of our interest, we suppose $E=0$.
The potential $U(R;0)$ is positive for $0<R<\rho_b$ (see Fig.~\ref{fig:potential}), 
and hence $\left(\frac{dR}{dt}\right)^2 < 0$ in that region. Thus there is no classical solution connecting $R=0$ and $R>\rho_b$. Then, as expected, the bubble nucleation is described by the quantum
tunneling through the potential wall in $0 \leq R \leq \rho_b$,
and after the nucleation the bubble expands. 
Note that the radius $\rho_b$, which is derived from the condition $U(R;E=0)$, 
is exactly the one that minimizes the Euclidean bounce action in the previous approach shown in Eq.~\eqref{bubble_radius}.
In order to evaluate the tunneling probability, we perform a Wick rotation $t\to -it_\mathrm{E}$ and obtain the Euclidean action
\begin{align}
S_{\mathrm{E}}[R]&=4\pi\sigma\int dt_\mathrm{E} \left(R^2\sqrt{f(R)+f(R)^{-1}\left(\frac{dR}{dt_\mathrm{E}}\right)^2}- \frac{R^3}{\rho_0}\right). \label{NGaction_Euclid}
\end{align}
The bounce solution for the Euclidean action satisfies the equation of motion with the flipped potential
\begin{align}
\left(\frac{dR}{dt_\mathrm{E}}\right)^2 - 2U(R;E=0)=0.
\end{align}
The tunneling rate is now evaluated with the bounce action that runs from $R=0$ to $\rho_b$ and then comes back to $R=0$~\cite{Coleman:1977py,Basu:1991ig}, 
which is calculated as
\begin{align}
B=2\times4\pi\sigma \int_{0}^{\rho_b} dR f^{-1} R^2\sqrt{1-R^2/\rho_b^2}=\frac{2\pi^2\sigma\rho_b^3} {(1+\rho_b/\rho_0)^2},
\label{bounce_action_result2}
\end{align}
by using Eq.~\eqref{integratedEOM_NG}.
The factor 2 comes from the bounce trajectory of $R: 0 \rightarrow \rho_b$ and 
$R: \rho_b \rightarrow 0$. 
One can see that the action obtained in (\ref{bounce_action_result2}) is indeed equivalent to \eqref{bounce_action_result1} obtained in the Euclidean scalar field theory with the thin-wall approximation. 
Therefore, as long as the thin-wall approximation is valid, where 
the dynamics of the bubble is characterized only by the size, tension, and bulk energy, 
it is enough to analyze it by the one-dimensional effective theory of bubble wall instead of the Euclidean scalar field theory.

\section{Vacuum decay in the Lorentzian path integral}
\label{sec_Lorentzian}
In this section, we revisit the vacuum decay process without relying on the Euclidean path integral. Instead, we adopt the Lorentzian path integral. Convergence of the path integral is guaranteed by the Picard-Lefschetz theory, which tells us how the contour of the integration should be deformed to make the path integral absolutely convergent.
See Appendix~\ref{PL_theory} for a brief review on the Picard-Lefschetz theory.
It is  difficult to perform the Lorentzian path integral in the full theory space 
even adopting the Picard-Lefschetz theory. So we adopt the one-dimensional model with a spherical thin-wall bubble
as in the previous section.

To perform the Lorentzian path integral, it is convenient to work in the canonical form of action whose kinetic term is quadratic in the conjugate momentum. 
Thus we will use the Polyakov-type action~\cite{Polyakov:1981rd,Duff:1987cs}\footnote{The path integral using Polyakov-type action is equivalent to that of NG-type action under the mean field approximation of a newly introduced Lagrange multiplier field (see Ref.\cite{Polyakov:1987}).}
\begin{align}
S_\mathrm{P}[X^\mu, \gamma_{ab}]=-\sigma\int_\mathcal{W} d^3x\frac12\sqrt{-\mathrm{det}\gamma_{ab}} \left(\gamma^{ab} h_{ab} (X^\mu)-1\right) +\Delta\!V\int_\mathcal{B} d^4\tilde x\sqrt{-\det g_{\mu\nu}},\label{Polyakovaction_wall}
\end{align}
where $\gamma_{ab}$ is the auxiliary induced metric on the bubble wall, which 
is classically equivalent to the NG-type action, investigated in the previous section. Indeed,  varying the action with respect to $\gamma_{ab}$ we obtain the following relation
\begin{align}
\gamma_{ab} = h_{ab} \label{onshell_metric},
\end{align}
which reproduces the original NG-type action after plugging it into~\eqref{Polyakovaction_wall}. Let us set $\gamma_{ab} = h_{ab}$ except for $\gamma_{00}$ to keep the off-shell contribution of the lapse function. Taking $\{ X^\mu(x) \} = \{ T(\tau), R(\tau), \theta, \phi \}$ again,
we have
\begin{align}
\gamma_{ab} dx^adx^b =-N^2(\tau)d\tau^2+R^2 (\tau) (d\theta^2+\sin^2\theta d\phi^2),
\end{align}
where $\sqrt{-\gamma_{00}} = N(\tau)$ is the lapse function. 
Consequently, one obtains the Polyakov-type action for which the dynamical variables are $R$ and $T$:
\begin{align}
S_\mathrm{P}[T,R,N]=4\pi\sigma\int^{\tau_1}_{\tau_0} d\tau \left\{\frac 1 2 R^2\left[N^{-1}(-f(R) \dot{T}^2 +f(R)^{-1} \dot{R}^2)-N \right] +\frac{R^3}{\rho_0} \dot{T}\right\} \label{Polyakovaction_R}.
\end{align}
Here and hereafter a dot denotes differentiation with respect to $\tau$.
One can easily check that NG-type action~\eqref{NGaction_R} is reproduced when the lapse function is on-shell:
\begin{align}
\frac{\delta S_\mathrm{P}}{\delta N}= 4\pi\sigma \frac 1 2R^2\left[ N^{-2}(f(R) \dot{T}^2-f(R)^{-1} \dot{R}^2)-1\right]=0\label{constraint}.
\end{align}

To compute the bubble nucleation probability based on the Lorentzian path integral, we here evaluate the transition amplitude from $R=R_0$ at $\tau=\tau_0$ to $R=R_1$ at $\tau=\tau_1$, where $\tau$ can be rescaled so that $\tau_0=0$ and $\tau_1=1$, since the transition amplitude is invariant under the affine rescaling of $\tau$. The transition amplitude is given by
\begin{align}
G(R_1;R_0)=\int_\mathcal{C} dN \int^{R(1)=R_1}_{R(0)=R_0}\mathcal{D}T\mathcal{D}R\exp{\left(iS_\mathrm{P}[T,R,N]\right)},
\label{propagator_wall}
\end{align}
where $N$ is 
a time-independent integration variable rather than a function after fixing the gauge of the time reparametrization invariance\footnote{The Faddeev-Popov ghost associated with the gauge fixing yields no contribution to the amplitude. See Refs.~\cite{Polyakov:1987,Mottola:1995sj}.} and $\mathcal{C}$, the integration range of $N$, denotes the contour along the real axis $(0+,\infty)$, to ensure proper time ordering \cite{Feldbrugge:2017kzv,Teitelboim:1983fh,DiazDorronsoro:2017hti}.
This expression has a simple interpretation as the summation of the transition amplitudes from $R_0$ to $R_1$ with various values of time elapsed. We choose $R_0=0$, as the bubble nucleation from nothing is of our interest.

The transition amplitude~\eqref{propagator_wall} could be dominated by the classical solutions for $(T,R,N)$ if exist, but eq.~\eqref{integratedEOM_NG} has no Lorentzian solution consistent with the boundary condition. 
Thus we adopt the technique in Ref.~\cite{Feldbrugge:2017kzv} here. 
That is,  we first perform the path integral with respect to the functions other than
the lapse function $N$, so that it is dominated by $N$-dependent  saddle solutions  obtained by taking the variation of the action with respect to
 $T$ and $R$ only.

Conservation of the Hamiltonian, $H$, and momentum conjugate to $T$ yields,
\begin{align}
H=&-4\pi\sigma \frac12R^2\left[ N^{-2}(f(R) \dot T^2-f(R)^{-1} \dot R^2)-1\right]=\mathrm{const.},\label{wall_hamiltonian_Polyakov} \\
&4\pi\sigma R^2f(R)\left(\frac{\dot{T}}{N} -\frac{R}{\rho_0f(R)}\right) =  E,\label{momentum_cons}
\end{align}
where $E$ is the conserved momentum conjugate to $T$.

Using Eq.~\eqref{momentum_cons} we can express ${\dot T}$ in terms of $R$ (and $E$) and 
obtain the equation of motion by taking the $\tau$ derivative of Eq.~\eqref{wall_hamiltonian_Polyakov}, 
whose explicit expression is given later. 
We require 
the boundary conditions as $R(0)=0, R(1)=R_1$, and the value of $H$
is determined by $E$, or equivalently, the boundary condition of ${\dot T}$. 
Since integration over $N$ is yet to be done, we do not require
the Hamiltonian to vanish here, unlike in \eqref{constraint}.

As a result, the transition amplitude is approximated as,
\begin{align}
G(R_1;R_0)&=\int_0^\infty dN A (N) \exp{\left(iS_\mathrm{eff}[N]\right)},\quad S_\mathrm{eff}[N]= S_\mathrm{P}[\overline{T},\overline{R},N],\label{propagator_eff}
\end{align}
where $\overline{T}$ and $ \overline{R}$ are the saddle solutions, and
$A(N)$ is a prefactor mainly arising from the path integral of the 
quantum fluctuations around the saddle solution. Here we do not need its exact form but only assume $A (N) \to 0$ in large $|N|$ limit. 
This assumption is supported by the fact that the kinetic term of $T$ (and $R$) is suppressed by $N$, though the analytic calculation is difficult to perform due to the non-trivial dependence on $R$ in the action. 

Since the transition amplitude~\eqref{propagator_eff} contains oscillatory integral which is difficult to evaluate, we calculate it using the Picard-Lefschetz theory \cite{Lefschetz:1975}, where the original integration contour $\mathcal{C}=(0+,\infty)$ along the real axis is replaced with the ``steepest descent'' contours for the real part of the exponent, $\mathrm{Re}[iS_\mathrm{eff}[N]]$~\cite{Feldbrugge:2017kzv}, passing through the saddle point $N = N_s$ on the complex $N$ plane. 
See Appendix \ref{PL_theory} for a brief review of this theory. 
The integration is dominated by the contributions around the critical points,
so that the transition amplitude is approximated as
\begin{equation}
G(R_1;R_0) \simeq \sum_s n_s  A_s \exp[i S_\mathrm{eff}(N_s)], \label{propagator_PL}
\end{equation}
where $A_s$ is the leading-order Gaussian integral around $N_s$, $n_s=\pm 1$ or $0$ 
counts the number of times that the deformed contour passes through each critical point, 
taking into account its orientation.

We must carefully examine if there are non-trivial contributions on the integral 
from the complex infinity in addition to the one from the saddle points.
In our model, the contour is deformed to the lower-half plane of the complex $N$, 
as we will see.
Since the effective action $S_\mathrm{eff}[N]$ grows as ${\cal O}(N)$ with a negative coefficient in the large $|N|$ limit, the contribution to the integral from the complex infinity in the lower-half plane
vanishes by virtue of the Jordan's lemma if the prefactor $A$ converges to zero at infinity. This argument holds true even when $S_\mathrm{eff}[N]$ behaves as a power of $N$, which is verified by just changing variable from $N$ to the power of $N$. See Ref. \cite{Feldbrugge:2017kzv} for further detailed discussions.

Now the problem reduces to looking for the saddle points of the effective action. Combining Eqs.~\eqref{wall_hamiltonian_Polyakov} and \eqref{momentum_cons}, one obtains
\begin{align}
\frac{\dot R^2}{N^2}+2V(R)=0,\quad \mathrm{with}\  2V(R)=\left(1-\frac{2H}{4\pi\sigma R^2}\right)f(R)-\frac{\left(E/4\pi\sigma+R^3/\rho_0\right)^2}{R^{4}}\label{integratedEOM_Polyakov},
\end{align}
where $\overline{R}$ is replaced with $R$ by abusing the notation. It is essential that here we can identify $E$ as the energy of a vacuum bubble~\eqref{bubble_energy}. To see this correspondence, let us remind ourselves of the bubble energy in the NG-type action, Eq.~\eqref{bubble_energy}. 

Imposing the relation
\begin{align}
N d\tau =\sqrt{f-\frac{1}{f}\left(\frac{dR}{dt}\right)^2} dt 
\end{align}
between $\tau$ and $t$,
the integrated equation of motion Eq.~\eqref{integratedEOM_NG} is rewritten as
\begin{align}
N^{-2}\left(\frac{dR}{d \tau}\right)^2 +f-\frac {\left(E/4\pi\sigma +R^3/\rho_0\right)^2}{R^4}=0.
\end{align}
Here we have used 
\begin{equation}
\frac{fR^2}{\sqrt{f - f^{-1} (dR/dt)^2}} = \frac{E}{4 \pi \sigma} + \frac{R^3}{\rho_0}, 
\end{equation}
which follows from Eq.~\eqref{bubble_energy}. This reproduces the equation of motion derived from the Polyakov-type  action~\eqref{integratedEOM_Polyakov} with the Hamiltonian constraint $H=0$. 
Therefore, the conjugate momentum $E$ can be regarded as the bubble energy in the Polyakov-type action. In the following discussion, we set it to be zero, since the bubble nucleation from vacuum is of our interest, while the Hamiltonian $H$ is treated as a parameter that characterizes the saddle solution.

To investigate the system with $E=0$, it is convenient to change the variables in Eq.~\eqref{integratedEOM_Polyakov} to $\Sigma=R^2$ as
\begin{align}
\frac{\dot\Sigma^2}{N^2}+2V_\Sigma(\Sigma)=0,\quad  \mathrm{with} \quad 2V_\Sigma(\Sigma)=4\left[-\frac{ \Sigma^2}{\rho_b^2}+\left(1+\frac{2H}{4\pi\sigma l^2}\right)\Sigma-\frac{2H}{4\pi\sigma }\right],\label{integratedEOM_polyakov_S}
\end{align}
so that it becomes the form for an inverted harmonic oscillator. The equation of motion is now obtained by taking the $\tau$ derivative of Eq.~\eqref{integratedEOM_polyakov_S} as
\begin{equation}
{\ddot \Sigma} = 2 N^2 \left[\frac{2 \Sigma}{\rho_b^2} - \left(1+\frac{2 H}{4 \pi \sigma l^2}\right)\right].\label{EOM_polyakov_S}
\end{equation}
The effective action can be expressed in terms of the solution of Eq.~\eqref{EOM_polyakov_S}, $\Sigma$, as
\begin{align}
S_\mathrm{eff}[N]=4\pi\sigma N\int_0^1d\tau\left(\frac{\Sigma^2}{\rho_0^2(1-\Sigma/l^2)}-\Sigma+\frac{H}{4\pi\sigma}\right).\label{effective_action}
\end{align}
In the following, we investigate the solution of the equation of motion (Eq.~\eqref{EOM_polyakov_S}) 
and perform the $N$ integration with the help of the Picard-Lefschetz theory for several cases.

\subsection{Nucleation of a critical vacuum bubble ($R=\rho_b$)}

First, we consider the nucleation of a critical bubble whose radius is $R= \rho_b$ at which the Euclidean bounce action (\ref{bounce_action_thin}) is minimized. 
It is the first check how the Lorentzian method works compared to the Euclidean method.
We solve the equation of motion (Eq.~\eqref{EOM_polyakov_S})
imposing the boundary condition $\Sigma(\tau = 0)=0$ and ${\dot \Sigma}(\tau=0)=N\sqrt{2H/\pi\sigma}$, 
which results from Eq.~\eqref{integratedEOM_polyakov_S} and the reality of $R$, $\Sigma (\tau)>0$.
$H$ is determined by requiring the solution to satisfy another boundary condition
 $\Sigma(\tau = 1)=\rho_b^2$.
Consequently, we obtain\footnote{There is another saddle solution, but it takes the value $\Sigma = l^2$
at some regions in $0<\tau<1$.  Thus  the effective action diverges, and hence we identify that it is irrelevant solution. }
\begin{align}
\Sigma{(\tau)}&=\frac{\rho_b^2}{\rho_0+\rho_b}\text{csch}^{2}\frac{N}{\rho_b}\sinh\frac{N \tau}{\rho_b}\left( \rho_b\sinh\frac {N \tau}{\rho_b} + \rho_0 \sinh\frac {N(2-\tau)}{\rho_b}\right)\\
H&=2\pi\sigma \frac{\rho_0^2\rho_b^2}{(\rho_0+\rho_b)^2}\coth^{2}\frac{N}{\rho_b}.\label{Hamiltonian}
\end{align}
Plugging them into Eq.~\eqref{effective_action}, 
the effective action is expressed as
\begin{align}
S_\mathrm{eff}[N]=\frac{2\pi\sigma \rho_b^3}{(1+ \rho_b/\rho_0)^2}\left(\coth\frac{N}{\rho_b}-\frac{N}{\rho_b}\right),\label{critical_action}
\end{align}
which can be analytically continued except for the singularities at $N \in i\pi \rho_b \mathbb{Z}$. 

Let us now deform the original integration contour in Eq.~\eqref{propagator_eff} to
the one passing through the relevant saddle point of the effective action
according to the recipe of the Picard-Lefschetz theory. 
The saddle points, or equivalently, critical points, and the corresponding values of the effective action on them are 
evaluated as
\begin{align}
&N_k = i\pi \rho_b (k+1/2), \quad k\in\mathbb{Z}\\
&S_\mathrm{eff}[N_k]=\frac{-i(2k+1)\pi^2\sigma \rho_b^3}{(1+\rho_b/\rho_0)^2}.
\end{align}
Note that there are infinite number of the saddle points due to the periodicity of the derivative of the action. 

\begin{figure}[htbp]
    \centering
    \includegraphics{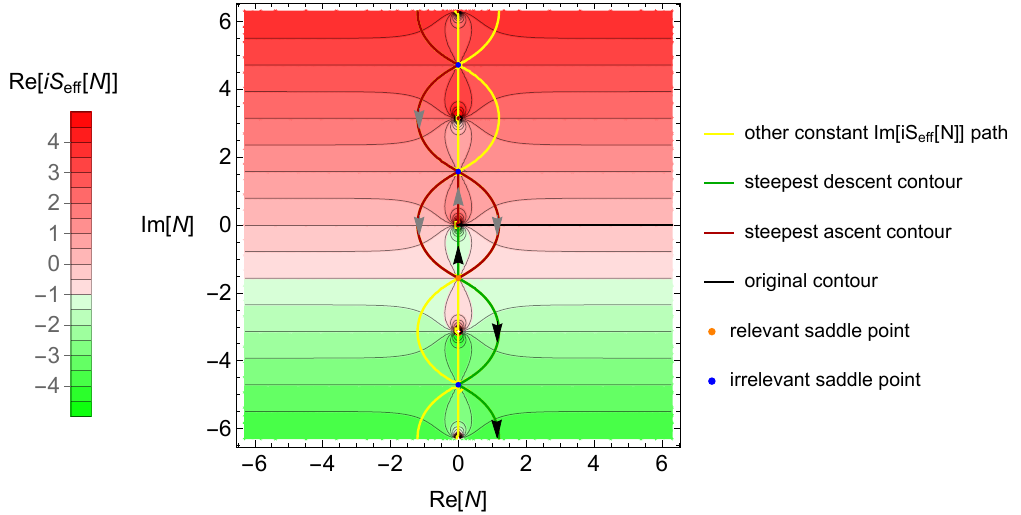}
    \hspace{5truemm}
    \caption{The value of $\mathrm{Re}[iS_\mathrm{eff}[N]]$ on the complex $N$ plane for the critical bubble nucleation is shown.
     In the red region $\mathrm{Re}[iS_\mathrm{eff}[N]]$ is larger than its value at the saddle point $N_{-1}$, indicated by the orange point, whereas in the green region it is smaller. 
We choose the parameters as
    $\rho_b=1,\ l=3$. $\sigma$ is chosen to rescale $\mathrm{Re}[iS_\mathrm{eff}[N_{-1}]]$ to $-1$.
          The green and red contours with arrows represent the steepest descent and ascent contours for the saddle point $N_{-1}$ after adding the infinitesimal perturbation $\delta S_\mathrm{eff}=+i \epsilon N$ to remove the degeneracy of the saddle points and steepest contours for the different saddle points, respectively.
     The direction of the arrows is the 
     decreasing one for $\mathrm{Re}[iS_\mathrm{eff}[N]]$.
     The steepest ascent contour
     intersects with the original contour (black line) only once. 
     Since other steepest ascent contours from other saddle points do not cross the original contour, we shall deform it to the steepest descent contours for $N_{-1}$. 
}
    \label{fig:ImSofN_critical}
\end{figure}

To determine how to deform the contour, we need to investigate the steepest descent
and ascent contours from each saddle point.
Note that the saddle points are degenerate with the order of 2, namely, $\partial S_\mathrm{eff}/\partial N =\partial^2 S_\mathrm{eff}/\partial N^2 =0$ at $N=N_k$, while the third derivative is non-vanishing there. 
This degeneracy can be removed 
by adding an infinitesimal perturbation, 
$\delta S_\mathrm{eff} = +i \epsilon N$.
We then find that the steepest descent 
contours from  $N_{-1}$, $\mathcal{J}_{-1}$,  reach the singularity at $N=0$ as well as the infinity ($N\rightarrow -i \infty$), with the latter passing near other saddle points $N_{k} (k<-1)$ and passing through the right side of the singularities.
The two steepest ascent contours from $N_{-1}$ go to $\mathrm{Im} N>0$ region 
with one of them crossing the original contour.
Since other steepest ascent contours from other saddle points do not cross the original integration contour and all the contours converge to the non-perturbed ones after taking $\epsilon \rightarrow 0$ limit, 
we conclude that according to the Picard-Lefschetz theory, 
we shall deform the contour to the steepest descent path from $N_-$ that connects 
$N=0$ to $N=-i \infty$ and close the contour by passing the infinity from $N=-i \infty$ to $N= + \infty$.
We show the real part of the exponent in the propagator, $\mathrm{Re}[iS_\mathrm{eff}[N]]$, behaviors of the steepest descent and ascent contours and the deformed contour in Fig.~\ref{fig:ImSofN_critical}. 
Note that the closed contour that connects the original and new contours does not 
contain any singularity in its inside, and the contribution from the complex infinity vanishes
since the effective action behaves as $S_{\rm eff} \sim - c N$ with $c>0$
in the large $|N|$ limit, which allows us to use Jordan's lemma to evaluate the Lorentzian path integral (\ref{propagator_eff}). 
The infinitesimal perturbation $+i\epsilon N$ also helps the integrand to converge at $N \rightarrow + \infty$. 
Thus the path integral is evaluated with Eq.~\eqref{propagator_PL} with $n_s=1$ for 
$s=-1$ whereas others to be zero\footnote{The contour also passes near the saddle points with $k<-1$. The contribution for the path integral around that point might not be suppressed so much compared to the one expected by the 
cubic approximation around $N_{-1}$, especially when we take $\epsilon \rightarrow 0$ limit 
for the infinitesimal perturbation. But such contributions  are 
nevertheless suppressed sufficiently if $\sigma \rho_b^3 \gg 1$.}.

Let us now calculate the propagator. By expanding
$iS_\mathrm{eff}[N]$ around the saddle point of $k=-1$ we obtain
\begin{align}
&iS_\mathrm{eff}[N]\simeq - \frac{\pi^2\sigma \rho_b^3}{(1+\rho_b/\rho_0)^2} -\frac {i 2\pi\sigma } {3(1+\rho_b/\rho_0)^2} \left(N-N_{-1}\right)^3.\label{around_saddle_critical}
\end{align}
The steepest descent contours that are chosen for the new contour of the path integral 
run in the direction satisfying $\arg[N-N_{-1} ]=\pi/2$ and $-\pi/6$. Thus the integral is approximated as the integration on the directions, which consequently yields
\begin{align}
\begin{split}
G(\rho_b;0)&\simeq \int_0^\infty  d x (-e^{i\pi/2}+e^{-i\pi/6})  \exp\left[- \frac{\pi^2\sigma \rho_b^3}{(1+\rho_b/\rho_0)^2}-\frac {2\pi\sigma } {3(1+\rho_b/\rho_0)^2} x^3\right] \\ 
&=\sqrt 3 e^{ -i\pi/3}\Gamma(4/3)\left(\frac{3(1+\rho_b/\rho_0)^2}{2\pi\sigma}\right)^{1/3}\exp\left[- \frac{\pi^2\sigma \rho_b^3}{(1+\rho_b/\rho_0)^2}\right],
\end{split}
\label{propagator_critical}
\end{align}
where we have taken $x \equiv |N-N_{-1}|$.
We can read off the exponent of the Lorentzian propagator, defined by $-L$, as 
\begin{equation}
L= \frac{\pi^2 \sigma \rho_b^3}{(1+\rho_b/\rho_0)^2}.
\end{equation}
After squaring the propagator to evaluate the transition rate, we obtain the same exponent as the Euclidean bounce action~\eqref{bounce_action_result2}
\begin{equation}
2L=B.
\end{equation}
This result is quite reasonable because the Hamiltonian \eqref{Hamiltonian} becomes zero at the saddle point on the imaginary axis, which means that the saddle point solution corresponds to the Euclidean bounce solution derived from the NG-type action. 
This result is a direct confirmation of the correctness
of the Euclidean formalism, because now it is derived from the Lorentzian path integral 
 taking into account the analytic structure of the action in the complex $N$ plane.

\subsection{Nucleation of a larger vacuum bubble ($R> \rho_b$)}

Next, we investigate the bubble nucleation whose size is different from the 
critical one. 
Note that it cannot be studied in the Euclidean formalism because an stationary 
configuration, which corresponds to the critical bubble nucleation is required in that formalism.
In the Lorentzian formalism, we do not require such a stationary configuration, and hence 
it becomes possible to study the bubble nucleation other than the critical bubble 
just by changing the boundary conditions for the path integral of $R$ and $T$, 
or equivalently, that for $\Sigma$ introduced in the previous section.

First, we consider nucleation of a larger bubble, 
which can be characterized by the boundary condition of $\Sigma$ for 
the equation of motion (Eq.~\eqref{EOM_polyakov_S}) as
$\Sigma (\tau = 1) / \rho_b^2 =1+p$ ($p>0$) to determine $H$ with other boundary conditions being kept the same. 
The general solution of~\eqref{EOM_polyakov_S} is given by
\begin{align}
\begin{split}
&\Sigma(\tau)=\sinh(z\tau)\left\{-\rho_b^2\left(1+\frac{v^2}{l^2}\right)\sinh(z\tau)+2\rho_b v \cosh (z\tau) \right\},\\
&\mathrm{with} \quad z=\frac{N}{\rho_b}, \quad v^2=\frac{H}{2\pi\sigma}. 
\label{general_solution}
\end{split}
\end{align}

$v$, or equivalently, $H$, is determined by requiring $\Sigma (\tau = 1) / \rho_b^2 =1+p$ as
\begin{align}
v=\frac{l^2}{\rho_b}\left(\cosh z-\frac{\rho_b}{\rho_0}F(z)\right)\text{csch} \, z,\quad  \mathrm{with} \quad  F(z)=\sqrt{\cosh z +\sqrt p\frac{\rho_0}{l}}\sqrt{\cosh z-\sqrt p\frac{\rho_0}{l}},\label{integration_constant}
\end{align}
where $F(z)$ is  analytically continued to the complex $z$ plane as far as possible with the standard branch of square root. 
Performing the integration in Eq.~\eqref{effective_action}, we obtain the effective action as
\begin{align}
S_\mathrm{eff}[N]&=\frac{4\pi\sigma l^2}{2\rho_b}\left[-\Sigma(\tau = 1)\coth z +2\rho_b v-(2l^2-\rho_b^2)z+\frac{\rho_b}{\rho_0}l^2\log\left(\frac{l^2\rho_0+(l^2-v\rho_0)\tanh z}{l^2\rho_0-(l^2+v\rho_0)\tanh z}\right)\right], 
\end{align}
and substituting $v$ given in Eq.~\eqref{integration_constant}, 
we yield the explicit form of the effective action as a function of $z$, 
\begin{align}
\begin{split}
S_\mathrm{eff}[N]=\frac{2\pi\sigma l^4}{\rho_b} &\left[\left(1+\frac{\rho_b^2}{\rho_0^2}\right)(\coth z-z)-p\frac{\rho_b^2}{l^2}\coth z \right.\\
&\left.-2\frac{\rho_b}{\rho_0}\left(F(z)\text{csch} z-\mathrm{arccoth}(F(z)\text{csch} \, z)\right)\right]. 
\label{general_action}
\end{split}
\end{align}
Note that it
reproduces the effective action for the critical bubble
~\eqref{critical_action} in $p\to 0$ limit. 

In order to perform the path integral by deforming the integration contour in the propagator~\eqref{propagator_eff}, we analytically continue the effective action to the complex plane.
Since the effective action includes the inverse hyperbolic function, $\mathrm{arccoth}$, 
we need to specify the branch appropriately. Here
we take the following branches, 
\begin{align}
\begin{split}
\mathrm{arccoth}(F(z)\text{csch} \, z)\to &
\begin{cases}
\mathrm{arctanh} (F(z)\text{csch} \, z)+i\pi/2 & \left((2k\pi\leq\mathrm{Im} z\leq(2k+1)\pi\right)\\
\mathrm{arctanh} (F(z)\text{csch} \, z)-i\pi/2 & \left((2k-1)\leq\mathrm{Im} z\leq 2k\pi\right)
\end{cases},\\
&\mathrm{with}\  \mathrm{arctanh} w =\frac12(\log(1+w)-\log(1-w)), \  k\in \mathbb{Z}
\end{split}
\end{align}
where we have taken the standard branch for log and square root. 
With this choice, we will see that the contour can be deformed to the one we can apply the 
Picard-Lefschetz theory.
We then obtain an analytic function defined on the whole complex $N$ plane except for the branch cuts located at
\begin{align}
z \in [i\left((k+1/2)\pi-\mathrm{arcsin} (\sqrt p \rho_0/l)\right),\  i\left((k+1/2)\pi+\mathrm{arcsin} (\sqrt p \rho_0/l)\right)].
\end{align}

\begin{figure}
    \centering
    \includegraphics{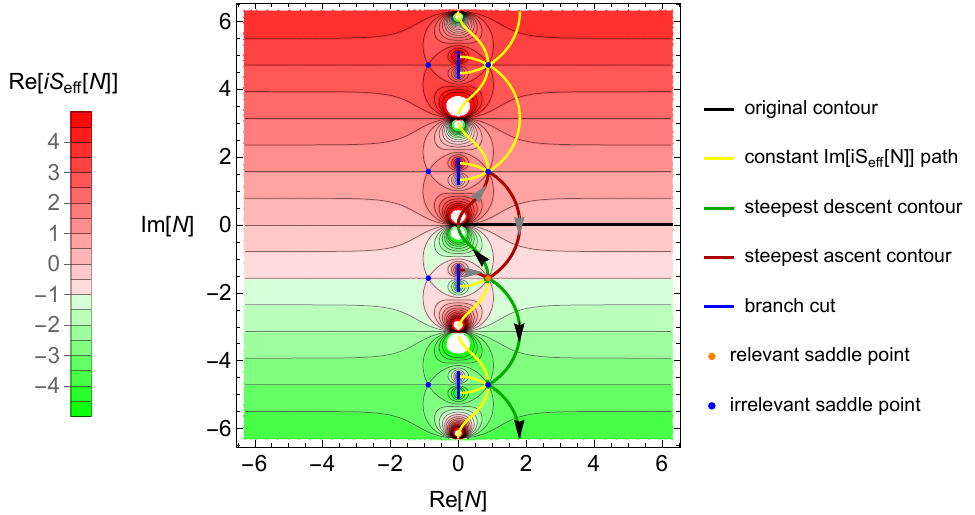}
    \hspace{5truemm}
    \caption{The value of $\mathrm{Re}[iS_\mathrm{eff}[N]]$ on the complex $N$ plane for the large bubble nucleation with parameters $p=1$. The blue lines show the brunch cuts and the other notations and parameter sets are the same as those in Fig. \ref{fig:ImSofN_critical}. 
    }
    \label{fig:ImSofN_large}
\end{figure}

As is done for the case of the critical bubble, we need to find the saddle points
and determine how to deform the contour for the path integral according to the 
Picard-Lefschetz theory. The derivative of the effective action
with respect to $N$ is calculated as
\begin{align}
\partial_N S_\mathrm{eff}[N]&=\frac{2\pi\sigma\rho_0 \rho_b^2(\cosh ^2z+p)^2\mathrm{csch}^{2}z}{\rho_0(1+\rho_b^2/\rho_0^2)\cosh z^2 -p\rho_0\rho_b^2/l^2+2\rho_b F(z)\cosh z },
\end{align}
from which we find the critical (or saddle) points are located at
\begin{align}
N_{\pm,k}=z_{\pm,k}/\rho_b,\ z_{\pm,k}=i(k+1/2)\pi \pm \log(\sqrt{p}+\sqrt{p+1}).
\end{align}
Similar to the case of the critical bubble, the saddle points are degenerate with the order of 2, 
but the degeneracy can be removed by a symmetry-breaking infinitesimal perturbation. 
We find there is only one saddle point whose steepest ascent contour crosses the original 
contour, $N_{+,-1}$. 
One of its steepest descent contour for $N_{+,-1}$ is connected to the origin and the other runs to  
$N \rightarrow -i \infty$. 
Thus, as before we shall deform the contour to the steepest descent contour of $N_{+,-1}$ 
followed by the contour that passes the infinity from $N=-i \infty$ to $N=+\infty$.
Note that the singularities are safely avoided in the contours.
The contour plot of $\mathrm{Re}[i S_\mathrm{eff}[N] ]$, is shown in Fig.~\ref{fig:ImSofN_large}.
Although the deformed contour, depicted with green line in Fig.~\ref{fig:ImSofN_large}, passes close to the saddle points of $k<-1$, 
their contributions are sufficiently suppressed
given that the semiclassical approximation is valid ($\sigma \rho_b^3 \gg 1$). 

The effective action on the saddle point $N_{+,-1}$ is evaluated as
\begin{align}
\begin{split}
S_\mathrm{eff}[N_{+,-1}]=&\frac{i
\pi^2\sigma\rho_b^3}{(1+\rho_b/\rho_0)^2}-\frac{2\pi\sigma\rho_0^2\rho_b^3}{(\rho_0^2-\rho_b^2)^2}\left[\sqrt{p(1+p)}(\rho_0^2-\rho_b^2)\right.\\
&\left.+(\rho_0^2+\rho_b^2)\mathrm{arctanh}\left(\sqrt {p/(p+1)}\right)-2\rho_0\rho_b\mathrm{arctanh}\frac{\sqrt {p/(p+1)}\rho_0}{\rho_b}\right], 
\end{split}\label{saddle_action_large}
\end{align}
and that around it is expanded as\footnote{In $p\to 0$ limit, the third order coefficient in the expansion does not coincide with that of critical bubble nucleation~\eqref{around_saddle_critical} due to the contribution from the branch cuts located near the saddle point (see Fig.~\ref{fig:ImSofN_large}).}
\begin{align}
    &iS_\mathrm{eff}[N]\simeq - \frac{\pi^2\sigma \rho_b^3}{(1+\rho_b/\rho_0)^2}+i\mathrm{Im}[iS[N_{+,-1}]] -\frac {i 2\pi\sigma } {3} \left(N-N_{+,-1}\right)^3.
\end{align}

Since the imaginary part of $iS_{\rm eff}[N]$ is stationary for the steepest direction of $\arg[N-N_{+,-1} ]=\pi/2$ or $-\pi/6$, the integral is approximated by the integration along these directions. Then the propagator for the nucleation of a larger bubble is
\begin{align}
\begin{split}
G(\sqrt{1+p}\rho_b;0)&\simeq \int_0^\infty  d x (-e^{i\pi/2}+e^{-i\pi/6})e^{i\mathrm{Im}[iS[N_{+,-1}]]}  \exp\left[- \frac{\pi^2\sigma \rho_b^3}{(1+\rho_b/\rho_0)^2}-\frac {2\pi\sigma } {3} x^3\right] \\ 
&=\sqrt 3 e^{i(\pi/3+\mathrm{Im}[iS[N_{+,-1}]])}\Gamma(4/3)\left(\frac{3}{2\pi\sigma}\right)^{1/3}\exp\left[- \frac{\pi^2\sigma \rho_b^3}{(1+\rho_b/\rho_0)^2}\right].
\end{split}
\label{propagator_large}
\end{align}

It is remarkable that the exponential suppression factor is the same as that for the critical bubble \eqref{propagator_critical}. The propagator has the additional phase factor including the $p$-dependence as one can see in \eqref{saddle_action_large}, which can be interpreted as the effect of classical evolution from $R=\rho_b$ to $R=\sqrt{1+p}\rho_b$.
 To see this, let us go back to the original theory with the thin-wall approximation~\eqref{NGaction_R} and analyze the transition amplitude from critical bubble configuration to the large bubble.
In the Lorentzian path integral the on-shell action dominates, whose value is obtained along with the equation of motion~\eqref{integratedEOM_NG} as,
\begin{align}
\begin{split}
S_\mathrm{NG}&=4\pi\sigma \int_{\rho_b}^{\sqrt{1+p}\rho_b} dR f^{-1}R^2\sqrt{R^2/\rho_b^2-1}\\
&=2\pi\sigma\rho_b^3\ell^2\left[-r_1\sqrt{r_1^2-1}+2\ell\sqrt{\ell^2-1}\ \mathrm{arctanh}\frac{\ell\sqrt{r_1^2-1}}{\sqrt{\ell^2-1}r_1}+(1-2\ell^2)\log(r_1+\sqrt{r_1^2-1})\right],
\end{split}
\label{s_ng_phase_comp}
\end{align}
with $\ell = l/\rho_b,\ r_1=\sqrt{1+p}$.  Substituting $l = l (\rho_b)$ of \eqref{bubble_radius} into (\ref{s_ng_phase_comp}), we obtain  exactly the same quantity as the real part of the effective action~\eqref{saddle_action_large}. 
Therefore, our analysis with the Lorentzian path integral directly shows that the large-bubble nucleation is dominated by the process consisting of the nucleation of critical bubble with radius $\rho_b$ and subsequent classical expansion of the bubble. Interestingly, in a recent work~\cite{Matsui:2021oio} a simple system of quantum mechanics has been investigated with the Lorentzian path integral, and it is found that the transition amplitude is consistent with the conventional WKB analysis of Schr\"{o}dinger equation. Our analysis is consistent with them.

In summary, We have shown that formation of a larger bubble proceeds through two steps, namely, nucleation of a critical bubble and its subsequent classical evolution, so that its formation probability is the same as that of a critical bubble. This cannot be analyzed by the Euclidean formalism alone, which inevitably leads to critical bubble formation.
It is worth noting that it becomes possible by using the Lorentzian analysis 
with the help of the Picard-Lefschetz theory, 
since the computation technique developed in this study to estimate the vacuum decay rate is more general than the traditional Euclidean instanton technique.

\subsection{Nucleation of a smaller vacuum bubble ($R< \rho_b$)}
\begin{figure}
    \centering
    \includegraphics{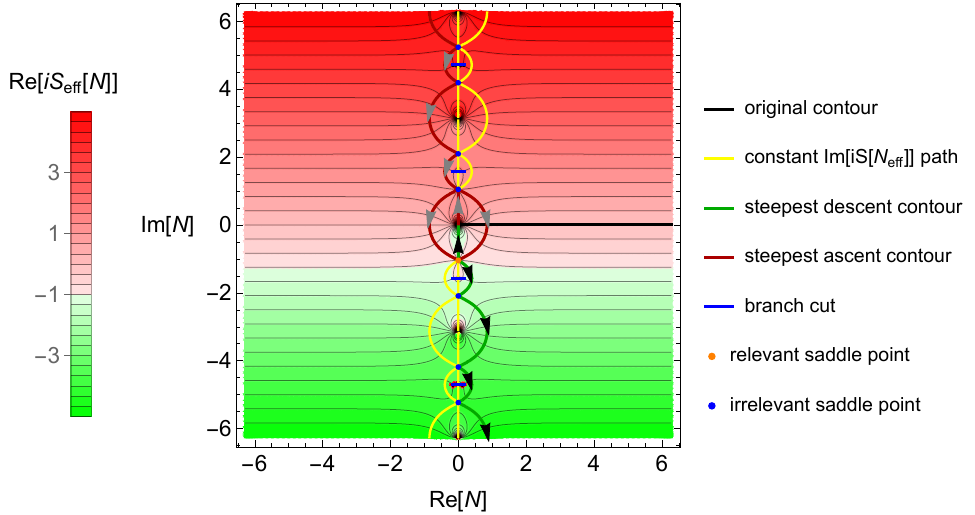}
    \hspace{5truemm}
    \caption{The value of $\mathrm{Re}[iS_\mathrm{eff}[N]]$ on the complex $N$ plane for the smaller bubble nucleation with parameters $p=-1/4$. The other notations and parameter sets are the same as those in FIG. \ref{fig:ImSofN_critical} and \ref{fig:ImSofN_large}.} 
    \label{fig:ImSofN_small}
\end{figure}

\begin{figure}[t]
\centering
    \includegraphics[width=0.7\textwidth]{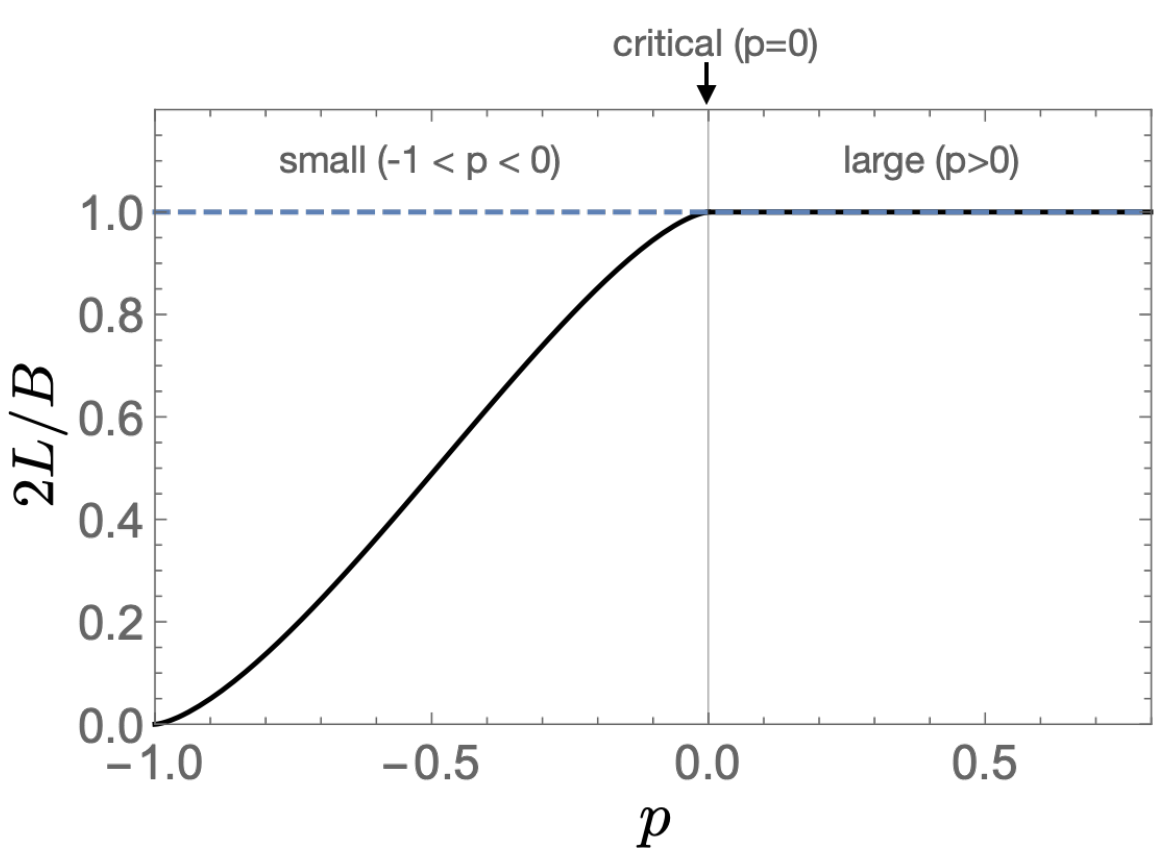}
\caption{A plot showing the exponent, $L (p)$, of the nucleation rates in (\ref{propagator_critical}), (\ref{propagator_large}) and (\ref{small_bubble_rate}). We here set $l/\rho_0=3$.
}
\label{fig:summary}
\end{figure}

Finally, we consider the nucleation of a smaller bubble with the boundary condition $\Sigma(\tau = 1)=(1+p)\rho_b^2$ for the equation of motion, similar to the 
larger bubble case, but
$-1<p<0$. 
Noting that we did not have to care about the value of $p$ when  we derived the solution of the 
equation of motion (on the real axis of $N$), the expression of the solution in this case is the 
same as the one obtained for the case of larger bubble. That is, we can just substitute a negative value 
of $p$ to the solution (Eq.~\eqref{general_solution}) with the Hamiltonian (Eq.~\eqref{integration_constant}) and also the effective action (Eq.~\eqref{general_action}).
However, the analytic continuation to the complex $N$ plane is slightly different from the case of a larger bubble.  Here we replace the choice of the branch as follows.
\begin{align}
&F(z)\to \tilde F(z)\equiv(-1)^k\sqrt{\cosh^2 z-p\rho_0^2/l^2},\\
&\mathrm{arccoth}(F(z)\text{csch} \, z) \to \mathrm{arccoth}(\tilde F(z)\text{csch} \, z)+k\pi,\\
&\quad\mathrm{with}\ (k-1/2)\pi\leq\mathrm{Im}[z]<(k+1/2)\pi,\ k\in\mathbb{Z}. 
\end{align}
Consequently, the branch cuts on the complex  plane are
located at
\begin{align}
z \in [i(k+1/2)\pi-\arcsin (\sqrt {-p} \rho_0/l),i(k+1/2)\pi+\arcsin (\sqrt {-p} \rho_0/l)], 
\end{align}
and the saddle points turned out to lie on the imaginary axis,
\begin{align}
N_{\pm,k}=z_{\pm,k}/\rho_b,\ z_{\pm,k}=i\left((k+1/2)\pi \pm \arcsin{\sqrt{-p}}\right).
\end{align}
We deform the path integral contour as depicted in $\mathrm{Re}[i S_\mathrm{eff}[N] ]$  shown in Fig.~\ref{fig:ImSofN_small}.
The value of the effective action is pure imaginary on the saddle points as
\begin{align}
&S_\mathrm{eff}[N_{\pm,k}]=\frac{-i(2k+1)\pi^2\sigma\rho_b^3}{(1+\rho_b/\rho_0)^2} \mp i C(p)\\
\begin{split}
&C(p)=\frac{2\pi\sigma\rho_0^2\rho_b^3}{(\rho_0^2-\rho_b^2)^2}\left[\sqrt{-p(1+p)}(\rho_0^2-\rho_b^2)\right.\\
&\quad\left.+(\rho_0^2+\rho_b^2)\arctan\left( \sqrt{-p/(1+p)}\right)-2\rho_0\rho_b\arctan\frac{\sqrt{-p/(1+p)}\rho_0}{\rho_b}\right]. 
\end{split}\label{saddle_action_small}
\end{align}
The second term, $C(p)$,  affects the bubble nucleation probability unlike the case of the large bubble nucleation. Since $C(p)$ is a decreasing function of $p$ with $C(0)=0$, we find 
$C(p)>0$ for
$-1<p<0$ regardless of $k$. 
The effective action around the relevant saddle point, $N_{+,-1}$, is expanded as
\begin{align}
    &iS_\mathrm{eff}[N]\simeq - \frac{\pi^2\sigma \rho_b^3}{(1+\rho_b/\rho_0)^2}+C(p) -\frac {i 2\pi\sigma } {3} \left(N-N_{+,-1}\right)^3. 
\end{align}
The steepest descent contour,
whose imaginary part is stationary, runs along the
directions of $\arg[N-N_{-1} ]=\pi/2$ or $-\pi/6$. 
Thus the propagator~\eqref{propagator_eff} is approximated by the following integral
\begin{align}
\begin{split}
G(\sqrt{1+p}\rho_b;0)&\simeq \int_0^\infty  d x (-e^{i\pi/2}+e^{-i\pi/6})  \exp\left[- \frac{\pi^2\sigma \rho_b^3}{(1+\rho_b/\rho_0)^2}+C(p)-\frac {2\pi\sigma } {3} x^3\right] \\ 
&=\sqrt 3 e^{i\pi/3}\Gamma(4/3)\left(\frac{3}{2\pi\sigma}\right)^{1/3}\exp\left[- \frac{\pi^2\sigma \rho_b^3}{(1+\rho_b/\rho_0)^2}+C(p)\right].
\end{split}
\label{small_bubble_rate}
\end{align}
After taking the square of this propagator, we obtain the suppression factor of the transition rate, 
which is twice the exponent of the Lorentzian propagator $L$ as 
\begin{equation}
2L(p)=\frac{2\pi^2 \sigma \rho_b^3}{(1+\rho_b/\rho_0)^2}-2C(p), 
\end{equation}
Note that it is smaller than the Euclidean bounce action $B$.
One can see that $2L(p) \to 0$ in the limit of $p \to -1$ and $2L(p) \to B$ for $p \to 0$. Since  $L(p)$ is a monotonic function (see Fig.~\ref{fig:summary}), the "nucleation rate" of a smaller bubble is higher than that of the critical one (\ref{propagator_critical}).
However, since nucleation of such a small vacuum bubble breaks the classical energy conservation law, this process is an off-shell phenomenon due to vacuum fluctuation.

\section{Conclusions and Discussions}
The standard procedure to evaluate the vacuum decay rate has traditionally been based on the construction of Euclidean instantons. 
However, it has been known that in some cases, there are ambiguities in evaluating it
because of the non-trivial analytic structure of the action of interest in the complex plane of time, 
which affects the way to perform the Wick rotation.
In this work, we have applied the Lorentzian path integral to compute the decay rate of a false vacuum state in a de Sitter and Minkowski background, as a first step to investigate the process
and establish the formalism to address the issue. To make the Lorentzian integral absolutely convergent, we have deformed the contour by following the Picard-Lefschetz theory. 
For simplicity, we reduced originally infinite degrees of freedom of a scalar field to  one dimension by using the thin-wall approximation. 
The dynamics of the bubble wall is described by the Polyakov-type action~\eqref{Polyakovaction_wall} that is quadratic in the bubble wall velocity. 

Based on the Lorentzian path integral, the nucleation process of a vacuum bubble was investigated for three cases: $R=\rho_b$, $R>\rho_b$, and $R< \rho_b$,  
with $\rho_b$ being the size of the critical bubble for which the Euclidean action is minimized. 
Note that the Euclidean formalism does not apply to
 the latter two cases. According to the 
Picard-Lefschetz theory,
we computed the Lorentzian path integral by analytically continuing the lapse function, $N$, to the complex plane and by evaluating the action near the relevant saddle point where the steepest descent and ascent contours intersect with each other. 

As a result, we have found that for the case of the critical bubble the transition probability led by the Lorentzian formalism~\eqref{propagator_large} is identical to the result based on the Euclidean formalism~\eqref{bounce_action_result2} and that a smaller bubble is nucleated with a higher probability which should be interpreted as quantum fluctuations without classical counterparts.  
The exponent of the exponential suppression factor is shown as a function of $p$ in Fig.\  \ref{fig:summary}. 
Our computation can be extended to the case of larger bubble, $R>\rho_b$, where the relevant saddle point is on the imaginary axis of $N$ (see Fig.~\ref{fig:ImSofN_large}). 
This is the reason why the Euclidean formalism is not applicable to this situation. 
In this sense, our computation has shown the advantage of the Lorentzian formalism and has demonstrated the consistency between the results obtained in the Lorentzian and Euclidean approaches without gravity. Recently, the real-time lattice simulation was performed to demonstrate the vacuum decay process in Ref.~\cite{Braden:2018tky}, and the nucleation rate of a vacuum bubble in the simulation was also in agreement with the instanton computation, which would support the validity of our method.

Note that in our approach of its current from, the one-loop corrections, {\it i.e.}, the prefactor of the vacuum decay rate, have not been fully evaluated since we have not taken all quantum fluctuations around the saddle points into account, including the degrees of freedom omitted when we adopt the thin-wall approximation\footnote{In Ref. \cite{Ai:2019fri}, the formula of the prefactor in the Lorentzian picture is provided, which is complimentary to our study,  and it matches with the prefactor obtained by Callan and Coleman in the Euclidean picture. It would be interesting to apply their formalism to our problem, but it is beyond the scope of the present study.}. We have not taken gravitational back reaction into account, either, as we have studied in a fixed background spacetime.
It will be an important direction to develop the Lorentzian approach so that one can compute the nucleation rate of vacuum bubbles fully incorporating gravity without relying on the Euclidean approach. Recently, the theory of vacuum decay catalyzed by cosmological impurities has been developed \cite{Steinhardt:1981ec,Hiscock:1987hn,Gregory:2013hja,Burda:2015yfa,Gregory:2018bdt,Mack:2018fny,Oshita:2018ptr,Oshita:2019jan,Koga:2019mee,Cuspinera:2019jwt,Gregory:2020hia,Hayashi:2020ocn,Gregory:2020cvy,Oshita:2020ksb,Firouzjahi:2020hfq,Koga:2020jok,Shkerin:2021zbf}. For a seed black hole, intriguingly, its Bekenstein-Hawking entropy \cite{Bekenstein:1972tm,Bekenstein:1973ur,Bekenstein:1974ax,Gibbons:1976ue,Hawking:1978jz} is involved in the nucleation rate of a catalyzed vacuum bubble. Reformulating this theory with the Lorentzian path integral will be another interesting direction, since the ambiguities in the Euclidean 
formalism are larger in such cases.

\acknowledgments
We thank Yusuke Yamada for useful discussions and comments. This work was partially supported by
JSPS Grant-in-Aid for JSPS Fellows 202114857(TH), Scientific Research (C) JP19K03842 (KK), Innovative Area 19H04610 (KK),  Research Activity Start-up 21K20371 (NO). 
TH is also supported by  Program of Excellence in Photon Science. 
NO is also supported by the FY2021 Incentive Research Project at RIKEN and the Special Postdoctoral Researcher (SPDR) Program at RIKEN.

\newpage
\appendix
\section{A short review of Picard-Lefschetz theory\label{PL_theory}}

In this appendix, we briefly review the Picard-Lefschetz theory applied to the simplest case with a single integral, following the discussion in Ref.~\cite{Feldbrugge:2017kzv}. Let us start from an oscillatory integral
\begin{align}
I=\int_0^\infty dN A(N) \exp\left(iS(N)\right),\label{PL_integral}
\end{align}
where $S(N)$ is a real-valued on the real axis of $N$ and is interpreted as a holomorphic function on the complex $N$ plane without any singularity in the relevant part of the domain. 
From the Cauchy's integral theorem, the integration can be performed by replacing the original integration contour $\mathcal{C}=(0,\infty)$ with a contour in the complex plane keeping both endpoints intact.
In the Picard-Lefschetz theory, the modified contour is chosen along 
the sum of the ``steepest descent'' contours where pass through the critical points $N_s$ 
where $\partial_N S(N)|_{N=N_s}=0$.
These points correspond to saddle points of
the real part of the exponent, namely the Morse function, $h(N) \equiv \mathrm{Re}[iS(N)]$, 
in the complex $N$ plane, 
and the steepest descent contours are the path along which $h(N)$ decreases at the rapidest.

To see this feature of the critical points, 
let us take advantages of the complex version of the Morse's lemma. 
It guarantees that $S(N)$ can be expanded around the critical point $N_s$ with an appropriate local complex coordinate $z$ that satisfies $N(z=0) = N_s$, as
\begin{align}
iS(N(z)) = iS(N_s)+z^2,
\end{align}
which leads to 
\begin{align}
h(N(z)) = h(N_s)+ \mathrm{Re}[z]^2-\mathrm{Im}[z]^2.
\end{align}
By looking at  
the Hessian matrix of $h(z)$ around the critical point, we find that $h(z)$ increases (decreases) most rapidly along  the $\mathrm{Re}[z]$  $ (\mathrm{Im}[z])$ direction. This means that
the critical point is a saddle point, with a steepest descent contour $\mathcal{J}_s$ for the $\mathrm{Im}[z]$ direction and a steepest ascent contour $\mathcal{K}_s$ for  the $\mathrm{Re}[z]$ direction.

For more detailed discussion on the properties of the steepest descent contour, let us introduce complex differential forms for $N=x+iy, N^*=x-iy$ as
\begin{align}
dN = d x+idy,\ dN^*=dx-idy,\ \partial_N=\frac 1 2\left(\partial_{x}-i\partial_{y} \right),\ \partial_{N^*}=\frac 1 2\left(\partial_{x}+i\partial_{y}\right).
\end{align}
In this case, we can introduce a parameter $\lambda$ originating 
from the saddle point along the steepest
descent contour, 
and with the Riemannian metric given on the complex plane,
\begin{align}
g_{ij}du^idu^j = dx^2+dy^2 = \frac 1 2\left(dNdN^*+dN^*dN\right), 
\end{align}
we define the Morse's gradient flow equation for downward flow,  
or the  steepest descent contour $\mathcal{J}_s$, 
\begin{align}
\frac{du^i}{d\lambda} = -g^{ij}\partial_{u^i}h.
\end{align}
Using this gradient flow equation,
we find that $h(N)$ monotonically decreases along the flow, 
\begin{align}
\frac{dh}{d\lambda} = \partial_{u^i} h \frac{du^i}{d\lambda}= - g^{ij}  \partial_{u^i} h \partial_{u^j} h< 0,
\end{align}
with the help of the positivity of the metric (This can be easily seen by taking $(u^1,u^2)=(x,y)$).
Moreover, for the imaginary part of the exponent $iS(N)$, $H(N)\equiv \mathrm{Im}[iS(N)]$,  
we find
\begin{align}
\frac{dH}{d\lambda} = \partial_{u^i} H(N)g^{ij} \partial_{u^j} h(N)=0, 
\end{align}
which can be derived from the Cauchy-Riemann equation for $iS(N)$. This means that
$H(N)$ is stationary along the steepest descent contour. In other words, the integrand in ~\eqref{PL_integral} is not oscillatory on the steepest descent contour $\mathcal{J}_s$. Since the absolute value of the integrand takes its maximum value at the saddle point and decrease 
to $-\infty$ along the flow originating from it, the integral along the flow is absolutely convergent. This contour is called a Lefschetz thimble.

Equivalently, the steepest ascent contour is also obtained by the upward flow equation
\begin{align}
\frac{du^i}{d\lambda} = + g^{ij}\partial_{u^i}h,
\end{align}
which leads to
\begin{align}
\frac{dh}{d\lambda}=(\partial_{u}h)^2 > 0,\quad \frac{dH}{d\lambda}=0,
\end{align}
and absolute value of the integrand takes minimum at the saddle point and blow up to $+\infty$ along the flow.

Let us then examine how the contour can be deformed to the one that pass through the 
critical points. In this case, the steepest contours, $\mathcal{J}_s$ and $\mathcal{K}_s$, emanating from the same saddle point 
intersects only once at that point. Note that with a suitable choice of orientation, this statement can be written with introducing 
the geometrical intersection number, $\braket{\cdot,\cdot}$, as
\begin{align}
\braket{\mathcal{J}_s,\mathcal{K}_{s'}} = \delta_{s,s'}. 
\end{align}
Based on the above consideration, we deform the contour smoothly to the sum of Lefschetz thimbles whose associated steepest ascent flow
crosses the original contour, by sliding the intersection point from the original contour to the saddle point.
The new contour is thus written as
\begin{align}
\mathcal{C_\mathrm{new}} = \sum_s n_s \mathcal{J}_s, 
\end{align}
where the neighboring thimbles are connected at $\lambda \rightarrow \infty$ where the amplitude of the
integrand is extremely suppressed. We also 
omit the path that connects both endpoints of the original contours to the end thimbles, 
which we expect to give no significant contributions to the integral.
Here $s$ runs all the critical points of $S(N)$ and the weight $n_s$ is identified as the intersection number of the original contour and the steepest ascent contour $\mathcal{K}_s$.  
Taking the orientation of the contour and the thimbles into account, $n_s = \braket{\mathcal{C},\mathcal{K}_s}$ takes the value $0, \pm1$.
Here we have used the fact that the smooth deformation from $\mathcal{C}$ to $\mathcal{C}_\mathrm{new}$ does not change the geometrical intersection number.

Since the phase of the integrand, $H(N)=\mathrm{Im}[iS(N)]$, is stationary on the deformed contour and the it is absolutely convergent as explained above, the absolute value of the integrand is mainly governed by its amplitude, $h(N)=\mathrm{Re}[iS(N)]$, around the saddle point. Therefore, we can easily evaluate the 
integral by the saddle point approximation in the complex $N$ plane as
\begin{equation}
I = \int_{\mathcal{C}_\mathrm{new}} dN A(N) e^{iS(N)} \simeq\sum_s n_s e^{iH(N_s)} \int_{\mathcal{J}_s} dN A(N) e^{h(N)}\simeq \sum_s n_s  A_s e^{iS(N_s)},  
\end{equation}
where $A_s$ is the leading-order Gaussian integral around the critical point $N_s$
with including the contributions from $A(N)$. 
Here contributions from the path connecting the both endpoints and thimbles are omitted, 
which is needed to be confirmed in the real calculations.

In the above discussion, we assumed that there are no degeneracy in the critical points and also
the steepest contours emanating from each critical point never coincide with each other. In our model, however, there are degenerate critical points $N_s$ where $S_\mathrm{eff}(N)=O((N-N_s)^3)$, as well as the degeneracy of the contours due to complex conjugation symmetry $N\to N^*$ of the effective action. In such cases, we can break these degeneracies by adding a complex perturbation $\epsilon(N)$ which breaks the symmetry in order to apply the Picard-Lefschetz theory.

\end{document}